\documentclass{ws-bormio02}

\begin{document}
\title{\Large What can be learned studying the distribution of 
the biggest fragment ?}
\author{{\bf E. Bonnet$^{1,2}$, F.~Gulminelli $^{3}$, B.~Borderie $^{1}$, 
N.~Le~Neindre$^{1}$, M.F.~Rivet $^{1}$}}
\address{
{\bf The INDRA and ALADIN Collaborations:}\\
$^{1}$Institut de Physique Nucl\'eaire, CNRS/IN2P3, Universit\'e Paris-Sud 11,
F-91406 Orsay-Cedex, France.\\
$^{2}$GANIL, DSM-CEA/IN2P3-CNRS, B.P.5027, F14076 Caen-Cedex, France, France.\\
$^{3}$LPC, IN2P3-CNRS, ENSICAEN et Universit\'e de Caen, F-14050 Caen-Cedex, France.\\
}
\maketitle
\abstracts{
In the canonical formalism of statistical physics, a signature of a first 
order phase transition for finite systems is the bimodal distribution of an 
order parameter. Previous thermodynamical studies of nuclear sources 
produced in heavy-ion collisions provide information which support the 
existence of a phase transition in those finite nuclear systems. Some results
suggest that the observable $Z_{1}$ (charge of the biggest fragment) can 
be considered as a reliable order parameter of the transition.
This talk will show how from peripheral collisions studied with the INDRA 
detector at GSI we can obtain this bimodal behaviour of $Z_{1}$.
Getting rid of the entrance channel effects and under the constraint of 
an equiprobable distribution of excitation energy ($E^{*}$), we use the 
canonical description of a phase transition to link this bimodal behaviour 
with the residual convexity of the entropy. Theoretical (with and without 
phase transition) and experimental $Z_{1}-E^{*}$ correlations are compared. 
This comparison allows us to rule out the case without transition. Moreover 
that quantitative conparison provides us with information about the
coexistence region in the $Z_{1}-E^{*}$ plane which is in good agreement 
with that obtained with the signal of abnormal fluctuations of 
configurational energy (microcanonical negative heat capacity).
}

\section{Introduction}
\indent
It is well known that a Liquid-Gas phase transition (PT) occurs in van der
Waals fluids. The similarity between inter-molecular and nuclear
interactions leads to a qualitatively similar equation of state which defines
the spinodal and coexistence zones of the phase diagram. 
That is why we expect a ``Liquid-Gas like'' PT for nuclei.
The order parameter is a scalar (one dimension) and is, in this case, the
density of the system (more precisely the density difference between the
ordered and disordered phase). The energy is also an order parameter because 
the PT has a latent heat.\\
When an homogeneous system enters the spinodal region of the phase diagram, 
its entropy exhibits a convex intruder along the order parameter(s) 
direction(s). The system becomes unstable and decomposes itself
in two phases. For finite systems, due to surface energy effects, we expect 
a residual convexity for the system entropy after the transition leading 
directly to a bimodal distribution (accumulation of statistics for
large and low values) of the order parameter. The challenge is to select an
observable connected to the theoretical order parameter of the transition, 
and to explore sufficiently the phase diagram to populate the coexistence 
region and its neighbourhood. Quasi-projectile sources
produced in (semi-)peripheral collisions cover a large range of dissipation
and consequently permit this sufficient exploration.\\
Several theoretical~\cite{Bot00,Gul05} and experimental
works~\cite{I51-Fra05,I54-Fra05,I61-Pic06} show that the biggest fragment has
a specific behaviour in the fragmentation process. In particular its size is
correlated to the excitation energy ($E^{*}$) of the sources. We can
reasonably explore whether the $Z_{1}$-$E^{*}$ experimental plane shows a
bimodal pattern. Other experimental signals obtained with multifragmentation 
data can be correlated with the presence of a phase transition in hot nuclei.
Indeed, abnormal fluctuations of configurational energy 
(AFCE)~\cite{ThNico,MDA00} can be related to the negative heat capacity 
signal~\cite{Cho99}, and the fossil signal of spinodal 
decomposition~\cite{I40-Tab03} can illustrate the density fluctuations 
occurring when the nuclei pass through the spinodal
zone~\cite{Cho04}. These two signals are not direct ones and need some 
hypotheses and/or high statistics. In this work we will present the study of 
the bimodality signal which is expected to be more robust and direct.
We will also show that its observation reinforces the conclusions extracted
from the two previous signals.

The idea is to show experimentally that the biggest fragment charge, $Z_1$,
can be a 
reliable observable to the order parameter of the PT. After an introduction 
of the canonical ensemble, we explain the procedure of
renormalization which allows to get rid of entrance channel and data sorting 
effects. Then, comparing experimental and canonical (E,Z1) distributions, 
we will show that the observed signal of bimodality
is related to the abnormal convexity of the entropy of the system. 
At the end, we propose a localisation of the coexistence zone deduced from 
a comparison between experimental data and the canonical description of a PT.

\section{Canonical description of first-order phase transition.}

Let us consider an observable E, known on average, free to fluctuate. The
least biased distribution will be a Boltzmann-Gibbs distribution 
(def.~\ref{Taylor1})~\cite{Balian}. If this observable is an order parameter 
of the system we have to distinguish two cases: with and without
phase transition.\\ 
\begin{eqnarray}
\label{Taylor1} \mathrm{P^{can}_{\beta}(E) = \frac{1}{Z^{can}_{\beta}}
e\;^{\;S(E) - \beta E} \; with \; Z^{can}_{\beta} = \int dE\; e\;^{\;S(E) -
\beta E}} \\
\label{Taylor2} \mathrm{S\left(E\right) \sim S\left(E_{\beta}\right) + 
\left(E-E_{\beta}\right) \frac{d}{dE}\;S \big|_{E_{\beta}}+
\frac{1}{2}\left(E-E_{\beta}\right)^{2} \frac{d^2}{dE^2}\;S 
\big|_{E_{\beta}}}\qquad	
\end{eqnarray}
For a one phase system (PT is not present), the microcanonical entropy, 
S(E)=log W(E) where W(E) is the number of microstates associated to the 
value of E, is concave everywhere. We can perform on it a saddle point 
approximation (eq.~\ref{Taylor2}) around the average value of $E$, 
$E_{\beta}$, meaning that the canonical distribution 
has a simple gaussian shape (eq.~\ref{Taylor3}). 
\begin{eqnarray}
\label{Taylor3} \mathrm{P^{(s.g.)}_{\beta}(E) = \frac{1}{\sqrt{2\pi
\sigma_{E}^{2}}}\;exp\left(-\frac{1}{2\sigma_{E}^{2}}\left(E-E_{\beta}\right)^{2}\right)
\; with \; \sigma^{2}_{E}=-\left(\frac{d^2}{dE^2}\;S \big|_{E_{\beta}}\right)^{-1}}\qquad \\
\label{Taylor4} \mathrm{P^{(s.g.)}(E^{*},Z_{1}) = \frac{1}{\sqrt{2\pi det\Sigma}}\:e^{\;-\frac{1}{2}\vec{x}\Sigma^{-1}\vec{x} } ,\;  \vec{x} = \left(
\begin{array}{ccc} E^{*}-E_{\beta} \\ Z_{1}-Z_{\beta} \end{array} \right) ,\; \Sigma = \left( \begin{array}{ccc} \sigma_{E}^{2} & \rho\;\sigma_{E}\sigma_{Z} \\ \rho\;\sigma_{E}\sigma_{Z} & \sigma_{Z}^{2} \end{array} \right)}
\end{eqnarray}
The parameters
of this gaussian are directly linked to the characteristics of the entropy.
In the same way we can define the minimum biased two dimensional 
distribution for the ($E^{*}$,Z1) observables leading to a
2D simple gaussian distribution~\cite{Gul07}
(def. \ref{Taylor4}). Parameters of this function gathered in the
variance-covariance matrix are also deduced from the curvature matrix 
of the 2D microcanonical entropy~\cite{Gul07,T41Bon06}.\\
\begin{eqnarray}
\label{Taylor5} \mathrm{P^{(d.g.)}(E^{*},Z_{1}) = N_{liq}\times P_{liq}^{(s.g.)}(E^{*},Z_{1}) + N_{gaz}\times P_{gaz}^{(s.g.)}(E^{*},Z_{1})}
\end{eqnarray}
When a system passes through a phase transition and enters in the spinodal 
region, the homogeneous system has a convex intruder in its microcanonical 
entropy along the order parameter(s) direction(s)~\cite{Gro02}. Instabilities occur and, 
due to the finite size of the system, the surface energy effects cause 
the non-additivity of the entropy leading at the end of the PT to a 
residual convex entropy for the two-phase system even at equilibrium. 
We cannot describe anymore the microcanonical entropy with a single saddle 
point approximation but we can introduce a double saddle point approximation. 
In this case the canonical distribution of the ($E^{*}$,Z1) observables can 
be described as the sum of two 2D simple gaussian distributions, one for 
each phase (def.~\ref{Taylor5})~\cite{Gul07,T41Bon06}.\\
 In the canonical ensemble, the energy distribution
$P^{can}_{\beta}(E^{*})$ as well as the two-dimensional distribution
$P^{can}_{\beta}(E^{*},Z1)$ are conditioned by the number of available states
$\exp{S}$ with a Boltzmann factor ponderation. The convex intruder in $S$ 
leads to a bimodality in the distribution~\cite{Gul07}. Experimentally, 
this relation is not so clear: the weight of the different states has no 
reason to be exponential and the measured distribution 
$P^{exp}_{\beta}(E^{*})$ is modified by a factor $g_{exp}(E^{*})$ which 
is determined in a large part by entrance channel effects
and data sorting : $P^{exp}(E^{*},Z_{1})=e^{S(E^{*},Z_{1})}g_{exp}(E^{*})$. The relative population of the different values of the
$E^{*}$ distribution looses its thermostatistic meaning 
($P^{exp}(E^{*}) \propto g_{exp}(E^{*}) P^{can}_{\beta}(E^{*})
e^{\beta E^*}$). We cannot therefore directly compare experimental and 
canonical distributions and deduce entropy properties of the system. 

\begin{eqnarray}
\label{renorm} \mathrm{P_{\omega}^{exp}(E^{*},Z_{1}) = \omega(E) \times P^{exp}(E^{*},Z_{1})}\\
\mathrm{with \; \omega(E^{*})=\left( \int P^{(exp)}(E^{*},Z_{1})\;dZ_{1} \right)^{-1}} \nonumber
\end{eqnarray}

\subsection{Renormalization method.}
In~\cite{Gul07}, a method was proposed to get rid of the experimental
effects. Assuming that the experimental bias $g_{exp}(E^{*})$ affects the 
$Z_1$ distribution only through its correlation with the deposited energy 
$E^{*}$ (phase space dominance),
 a renormalization of the ($E^{*}$,$Z_1$) distribution under the constraint 
 of an equiprobable distribution of $E^{*}$ (eq.~\ref{renorm}) allows to be 
 $E^*$-shape independent. If the system passes through a PT and the 
 correlation between $E^{*}$ and $Z_1$ is not a one-to-one correspondence, 
 it could reflect a residual convex intruder of the entropy.

\subsection{Spurious bimodality}
In principle one can ask whether the renormalization procedure given by 
eq.~\ref{renorm} can create spurious bimodality.
This does not seem to be the case for different schematic 
models~\cite{Gul07} but cannot be excluded \emph{a priori}. Another ambiguity 
arises from the fact that a physical bimodality can be hidden by the 
renormalization procedure if the correlation between $Z_1$ and $E^{*}$ is 
too strong. Bimodality can be also difficult to spot if the energy 
interval is not wide enough. For these reasons
 in the following we will compare the two canonical cases (with and without
transition) with the experimental distribution,
to check the validity of the obtained signal. 

\begin{figure}[!hbt]
\begin{center}
\includegraphics[scale=0.50]{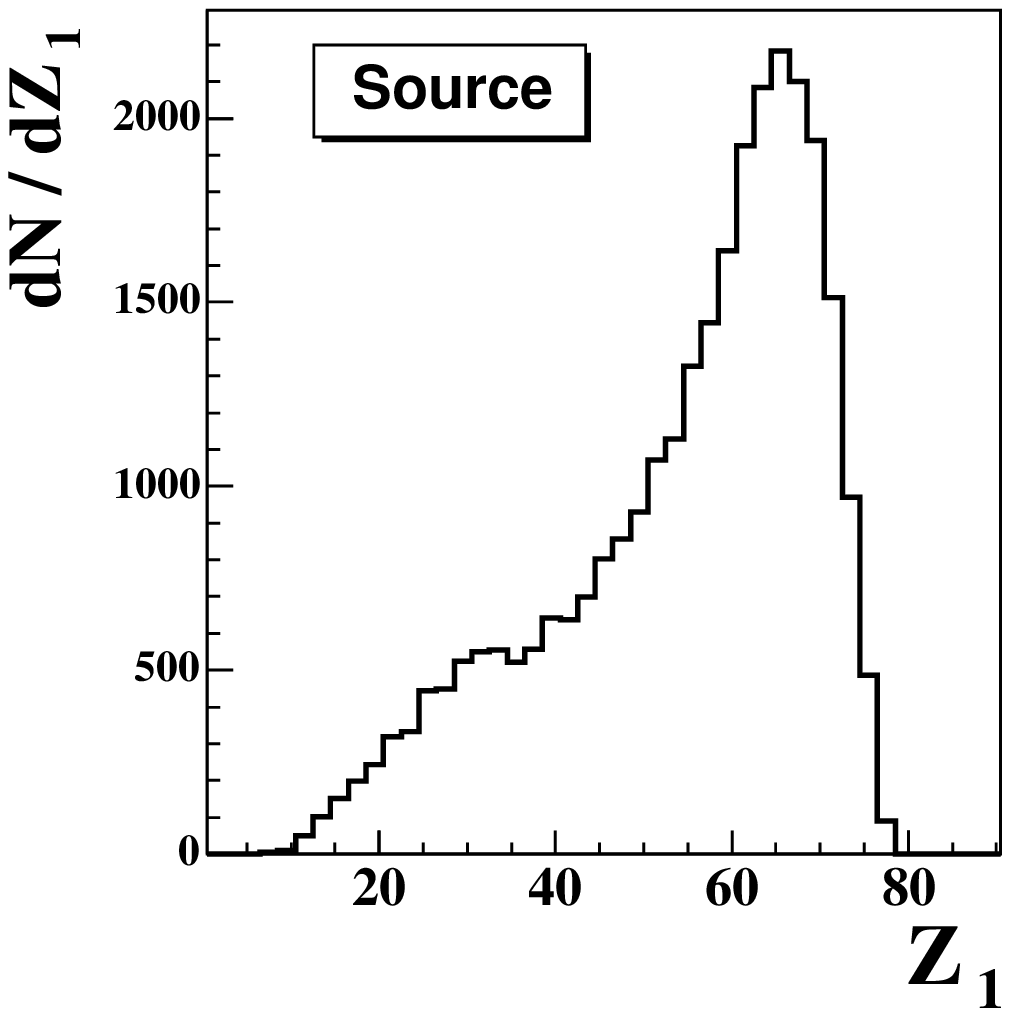}
\includegraphics[scale=0.50]{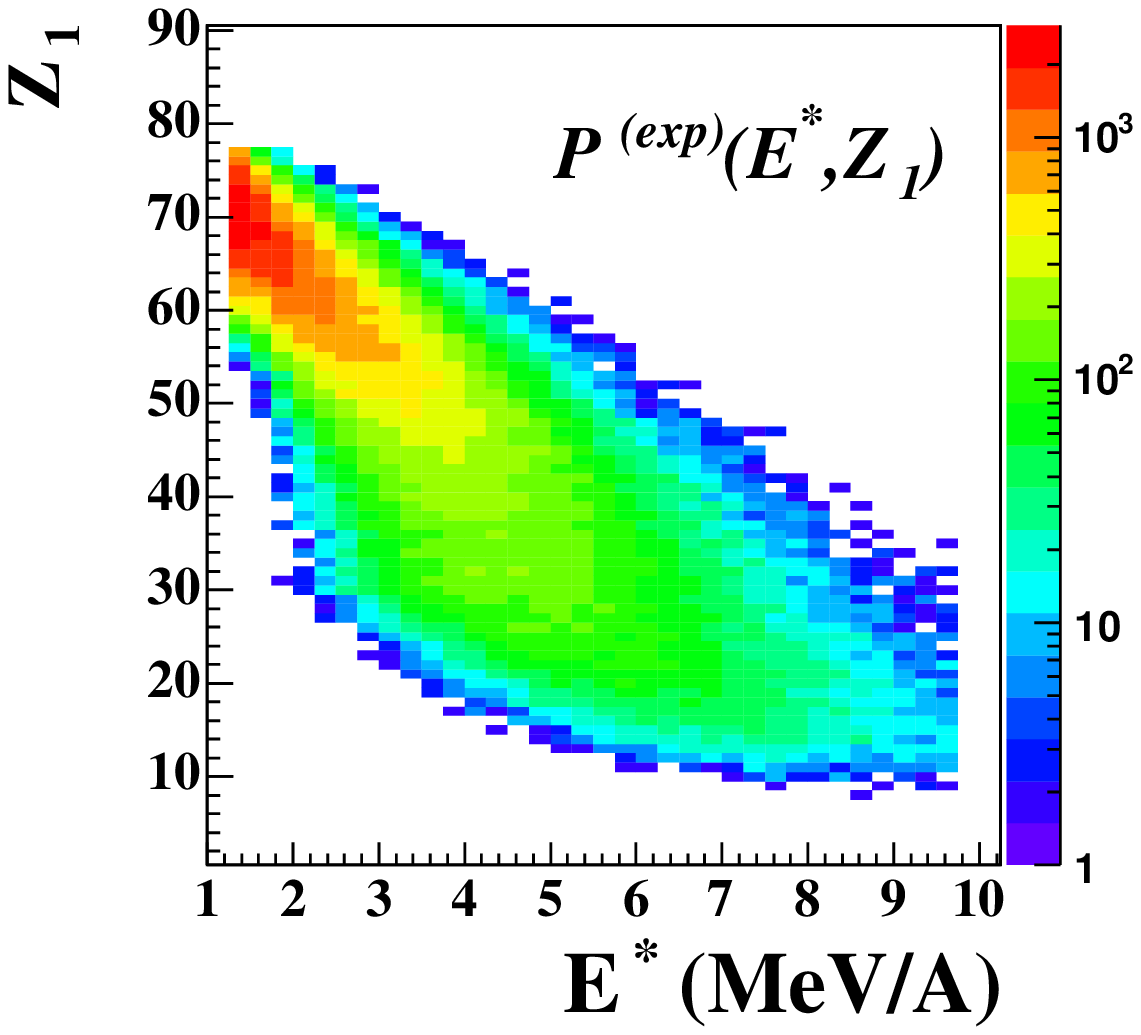}
\includegraphics[scale=0.50]{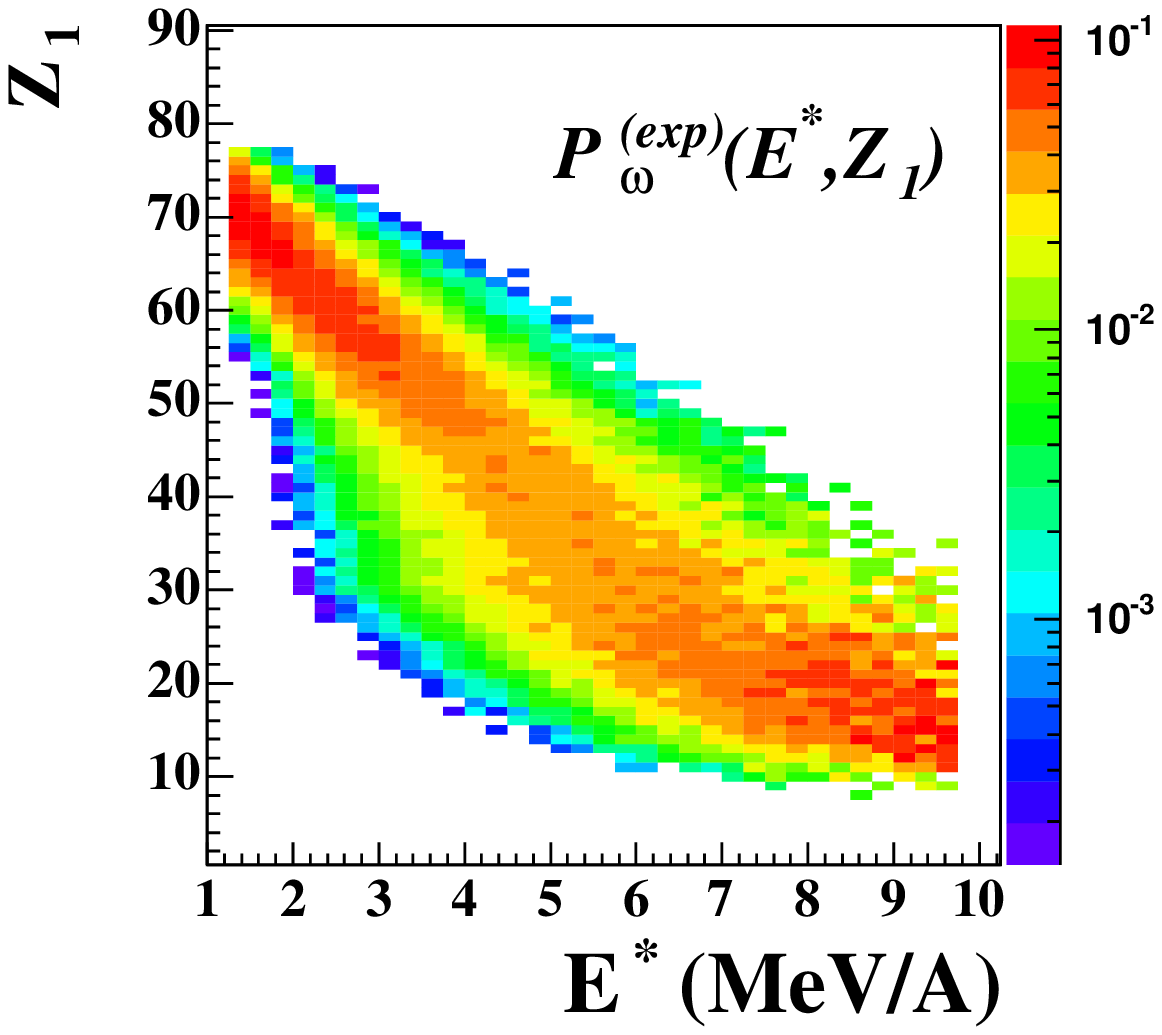}
\includegraphics[scale=0.50]{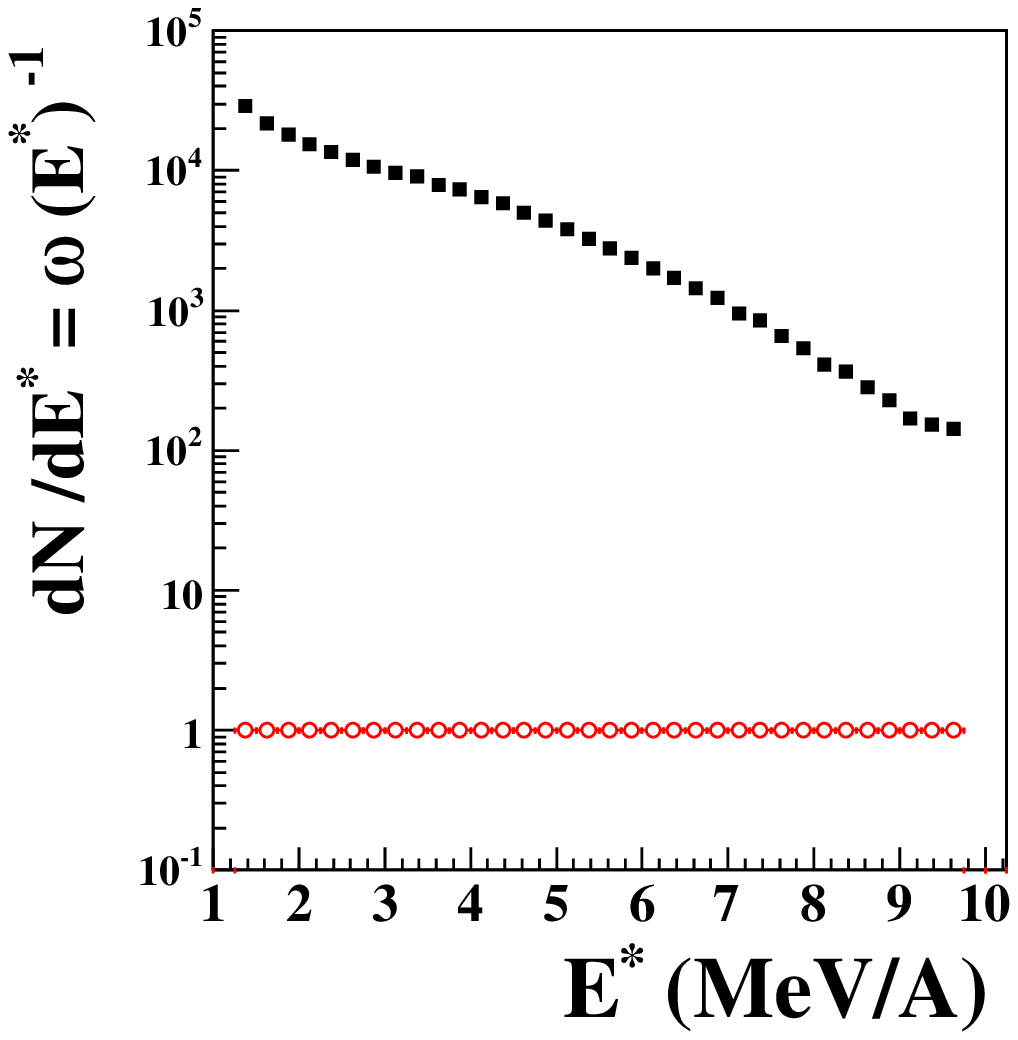}
\caption{\it Upper part : left : experimental distribution of the argest 
size fragment ($Z_1$) of source events; right : experimental correlation 
between $Z_1$ and the excitation energy ($E^*$). Lower part : left : 
experimental reweighted correlation between $Z_1$ and the excitation energy; 
right : excitation energy ($E^*$) experimental distribution of source events 
in black squares; the open red circles show this distribution after the 
renormalization process. For this, we keep only $E^*$ bins with a statistics 
greater than 100.}
\label{DistriZ1deb}
\end{center}
\end{figure}

\section{Data selection and first observation of $Z_{1}$ distributions.}
\indent
Data used in this present work are 80 MeV/A Au+Au reactions performed at 
the GSI facility and detected with the INDRA 4$\pi$ multidetector. We 
focus on peripheral and semi-peripheral collisions to study 
quasi-projectile sources (forward part of each event). To perform 
thermostatistical analyses, we select a set of events with a dynamically
compact configuration for fragments, to reject dynamical events which are 
always present in heavy-ion reactions at intermediate energies. 
We require in addition a constant size of the sources to avoid size evolution 
effects in the bimodality signal~\cite{T41Bon06,Bonnet07}. We evaluate 
the excitation energy using a standard calorimetry 
procedure~\cite{Cus93,I32-Vie02}. We compute the energy balance 
event-by-event in the centre of mass of the QP sources calculated with 
fragments only to minimize the effect of pre-equilibrium particles. 
Afterwards we keep only particles emitted in the forward part
of the QP sources and double their contribution, assuming an isotropic 
emission. In figure~\ref{DistriZ1deb} information on the experimental 
$Z_1$ and $E^*$ observables is shown, the latter covering a range between
roughly 1 and 8~MeV/A (lower-right part). Spinodal zone limits obtained with
the AFCE signal are around 2.5 and 5.8~MeV/A for this set of 
data~\cite{T41Bon06}.
 The shape of the distribution $P^{exp}(E^{*},Z_1)$ (upper right part) 
 shows the dominance of low dissipation-large $Z_1$ events and reflects 
 the cross-section distribution and data selection.
If we look at the corresponding $Z_1$ distribution (upper left part) we do 
not see any clear signal of bimodality: a large part of statistics is 
around 65-70, and only a shoulder is visible around 30-40. 
This particular shape could reflect the lack of statistics for the 
"gas-like" events. If we apply the renormalization procedure 
(eq.~\ref{renorm}) we obtain (lower left part) a $P_{\omega}^{exp}(E,Z_1)$ 
distribution which has a double humped shape, tending to prove that 
this procedure can reveal bimodality.

\begin{figure}[!hbt]
\begin{center}
\includegraphics[scale=0.50]{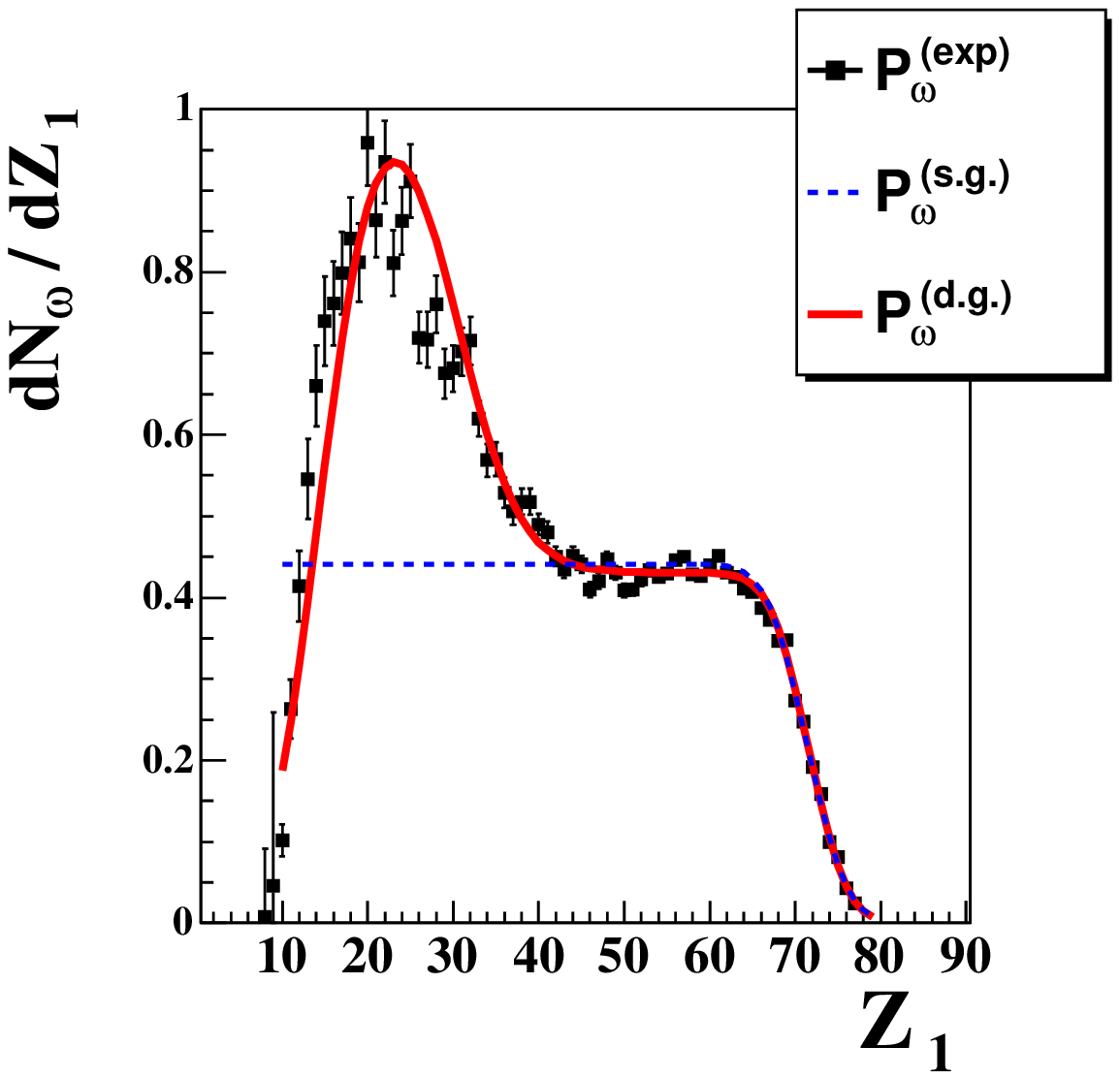}
\includegraphics[scale=0.50]{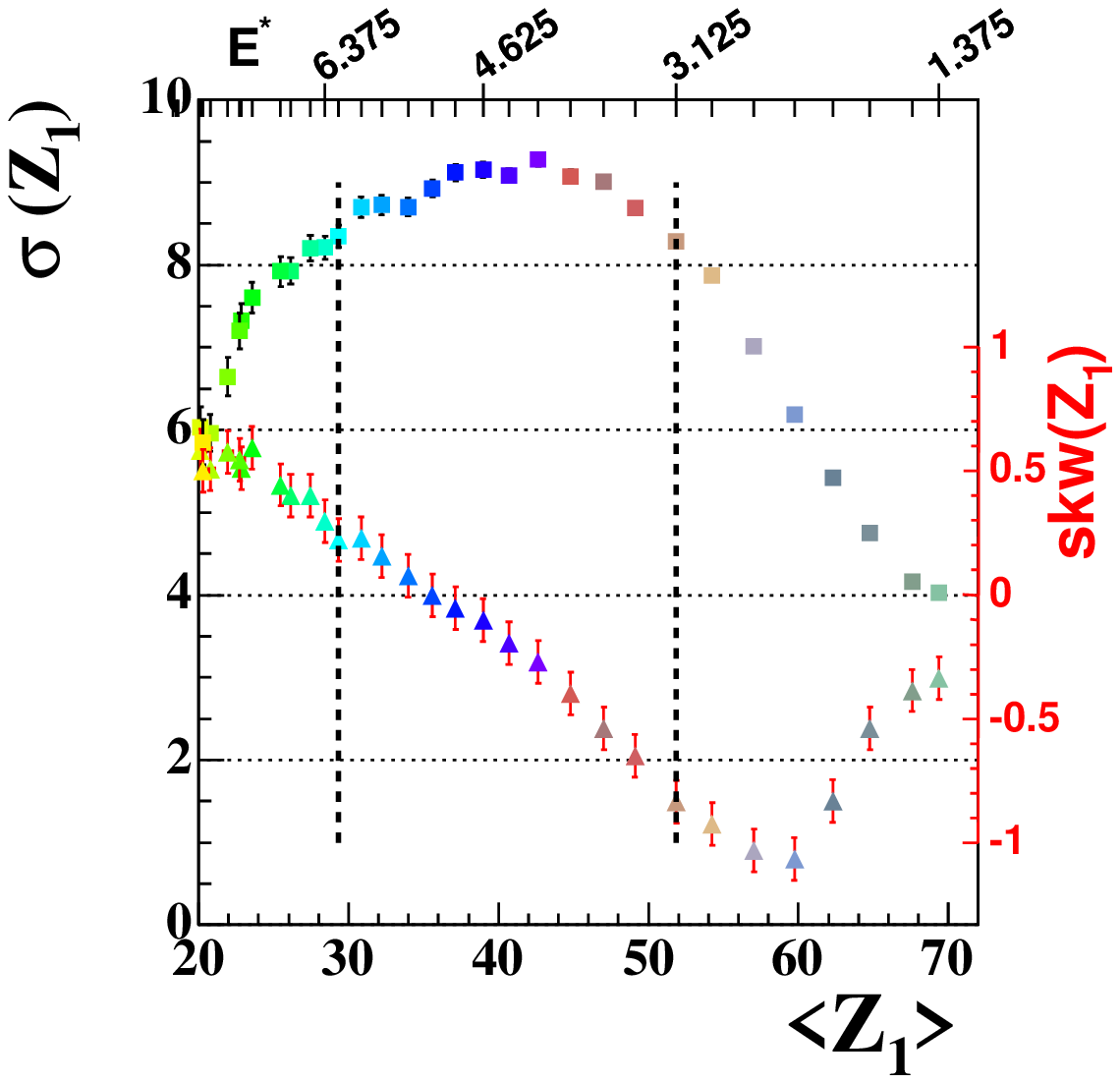}
\includegraphics[scale=0.50]{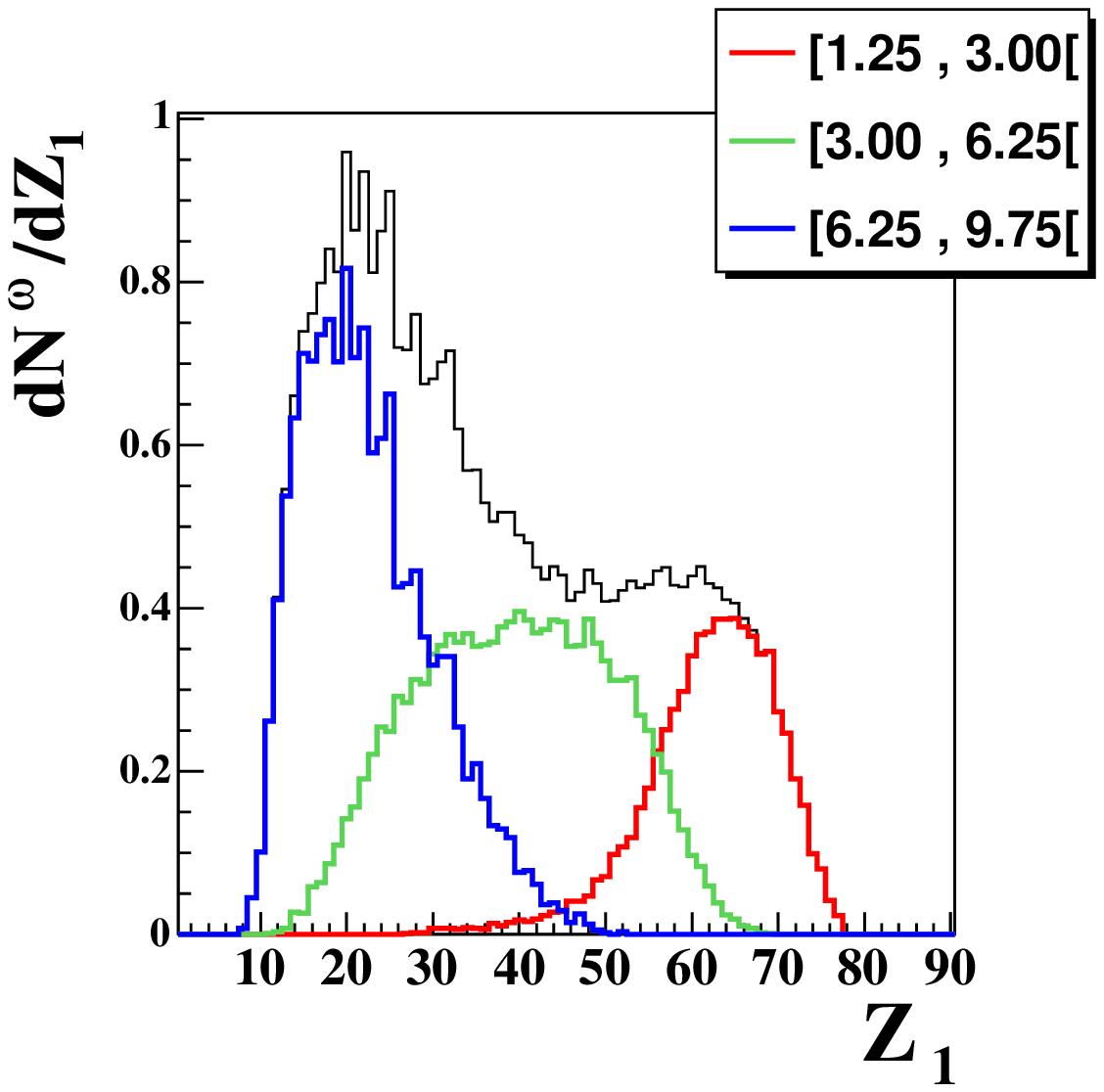}
\includegraphics[scale=0.50]{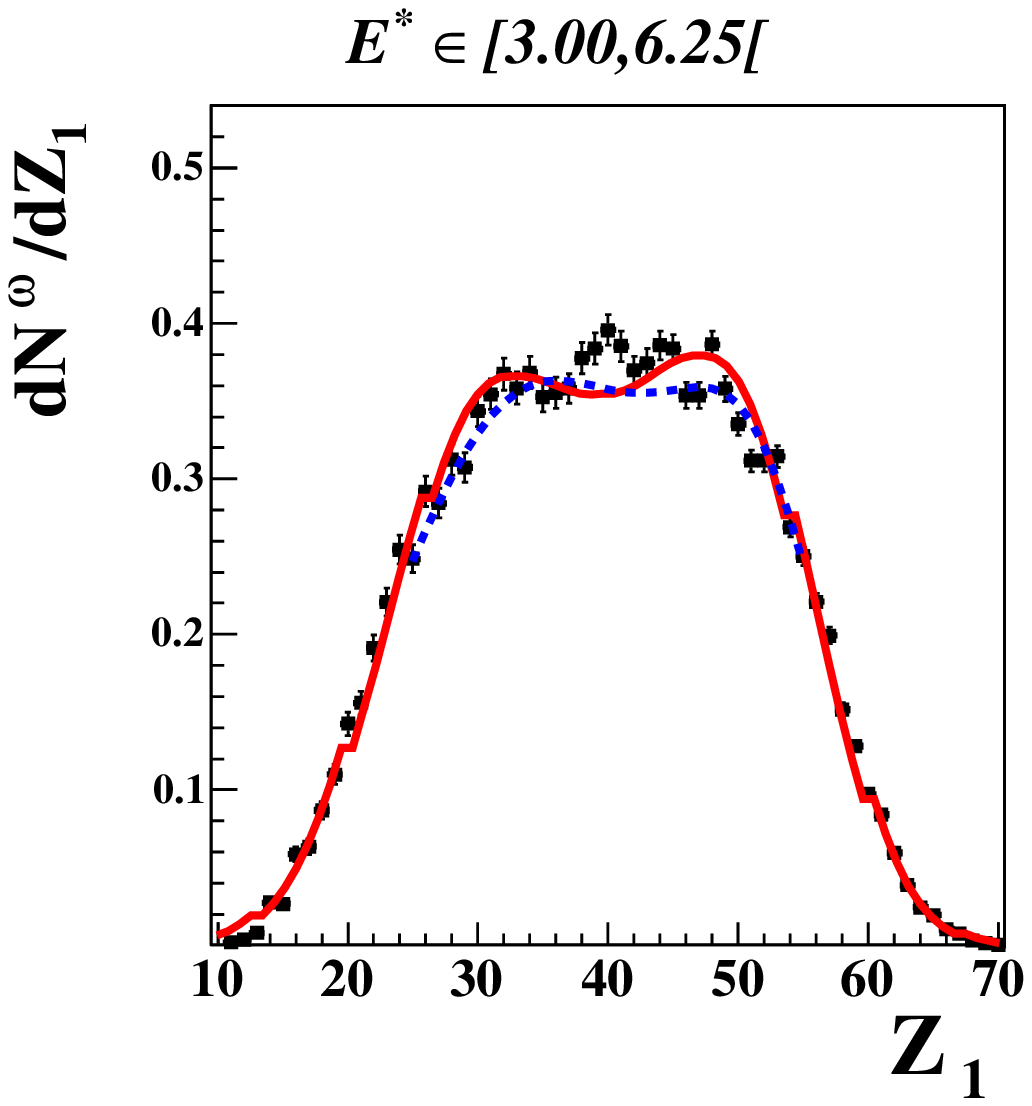}
\caption{\it Upper part: left: Largest size fragment ($Z_1$) experimental 
reweighted distribution (black squares with error bars); the blue dashed 
curve corresponds to the best solution obtained by comparing 
data and a single gaussian function (concave entropy, no PT), the red 
continuous curve corresponds to the best solution obtained by 
comparing data and a double gaussian function (convex entropy, PT); 
right: microcanonical sampling (fixed $E^*$) of the mean, the RMS and 
the skewness of the $Z_1$ distributions. For each bin of $E^*$ (upper X 
axis), RMS (colored squares-left Y axis) and skewness (colored 
triangles-right Y axis) are plotted as a function of the mean value 
($Z_1$-lower X axis); the two vertical dashed lines delimit the evaluated 
experimental spinodal zone where a quantitative comparison between data 
and PT case is performed. Lower part: left: same reweighted distribution 
of $Z_1$ as above (black curve); the three other distributions correspond 
to the three regions delimited by the vertical dashed lines 
(from left to right E$\in$[1.25,3.00[,[3.00,6.25[ and [6.25,9.75[); right:
best solutions obtained after the 2D comparison between data and canonical 
PT case; results are plotted for the $Z_1$ axis projection; 
the two solutions correspond to two different ranges of $Z_1$ where fits 
have been performed Z1$\in$[25,55] (dashed curve) and Z1$\in$[10,79] 
(continuous curve). The corresponding parameter values are listed in 
table~\ref{Resu_Fit2D}.}
\label{DistriZ1fin}
\end{center}
\end{figure}

\begin{table}[!hbt]
\begin{center}
\begin{tabular}{|c|c|c|c|}
\hline 	Parameters							&[10,79]  	& 	[25,55]	& Errors (\%) \\
\hline $\mathrm{\bar{E^{*}}_{liq}}$  	&  2,10 	 	& 	1,67 	  	& 23	\\
\hline $\mathrm{\sigma_{E^{*}_{liq}}}$	&  2,09 	 	& 	1,66 	  	& 23 	\\
\hline $\mathrm{\bar{Z_{1}}^{liq}}$  	&  60,2 	 	& 	62,2 	  	& 3 	\\
\hline $\mathrm{\sigma_{Z_{1}^{liq}}}$	&  12,9 	 	& 	9,85 	  	& 4 	\\
\hline $\mathrm{\bar{E^{*}}_{gaz}}$  	&  7,11 	 	& 	6,81 	  	& 4 	\\
\hline $\mathrm{\sigma_{E^{*}_{gaz}}}$	&  3,07 	 	& 	2,97 	  	& 3	\\
\hline $\mathrm{\bar{Z_{1}}^{gaz}}$  	&  21,1 	 	& 	23,8 	  	& 12 	\\
\hline $\mathrm{\sigma_{Z_{1}^{gaz}}}$	&  15,2 	 	& 	18,8 	  	& 2 	\\
\hline $\mathrm{\rho}$						&  -0,906 	& 	-0,860  	& 4 	\\
\hline $\mathrm{N_{liq}/N_{gaz}}$ 		&  1,12 	 	& 	0,66		& 52 	\\
\hline $\mathrm{N_{dof}}$					&  605	 	& 	387		&  -	\\
\hline $\mathrm{\chi^{2}}$					&  1488		& 	646		&  -	\\
\hline $\mathrm{\chi^{2}/N_{dof}}$		&	2,459	 	& 	1,669		&  - 	\\
\hline  
\end{tabular}
\caption{ \it Parameters values for the two best reproductions of data by 
the double gaussian function $\mathrm{P^{(d.g.)}_{\omega}}$ for two ranges 
in $Z_1$ [10,79] (first column) et [25,55] (second column); the third column 
gives relative errors computed with the previous values; 
$\bar{E^{*}}$,$\bar{Z_{1}}$,$\sigma_{E^{*}}$,$\sigma_{Z_{1}}$ stand 
respectively for centroids and RMS in the two directions ($E,Z_1$) of each 
phase (liquid and gas). The $\rho$ parameter is the correlation factor 
]-1,1[ between $Z_1$ and $E^*$ and the ratio $N_{liq}/N_{gaz}$ indicates
the repartition of statistics between the two phases. The three last lines 
give the number of degrees of freedom and the absolute and normalized 
$\chi^{2}$ estimator values.}
\label{Resu_Fit2D}
\end{center}
\end{table}

\section{Canonical-Experimental comparisons.}

To confirm that the two-hump distribution of $Z_1$ signals a convex 
intruder in the underlying entropy, in this section we compare  
the experimental reweighted distributions with the analytic expectation
for a system exhibiting or not a first order PT. We apply the same 
renormalization to $P^{(s.g.)}_{\beta}(E)$ and $P^{(d.g.)}_{\beta}(E)$ 
and try to reproduce the data. We focus on the projection on the $Z_1$ axis 
to perform the fit. The results are shown in the upper left part of 
fig.~\ref{DistriZ1fin}. The scatter points with errors bars correspond 
to the data; the continuous (respectively dashed) curve corresponds to 
the best solution obtained for the double (respectively simple) reweighted 
gaussian. We can clearly distinguish the two behaviours, the no-transition 
case can not curve itself in the Z1=40-50 region and can only reproduce 
one phase. The fact that data are reproduced with the functional describing 
a first order transition allows us to associate the experimental bimodality 
signal to a genuine convexity of the system entropy. This confirms also 
that the $Z_1$ observable is linked to the order parameter of the 
transition. To obtain more quantitative information we have to better 
localize the spinodal region. To do this, we look at the second and third 
moments of the $Z_1$ distribution for each bin of $E^*$. Their evolution 
is plotted on the upper right part of fig.~\ref{DistriZ1fin} as a
function of the mean value of $Z_1$ (lower X axis) and $E^*$ (upper X 
axis). The squares (left Y axis) stand for the sigma ($\sigma$) of 
the distribution and the triangles (right Y axis) for the skewness (skw). 
$\sigma$ shows a maximum in the range 30-40 for $<Z_1>$. This maximum 
of fluctuations signs the core of the spinodal zone which corresponds 
to the hole in $P_{\omega}^{exp}(E,Z1)$ distribution. All values of $Z_1$, 
for a given $E^*$, are more or less equivalent. In the same region the 
skewness changes sign, illustrating the change in the distribution 
of asymmetry, with a value close to zero when the distribution approaches 
a normal one. The two vertical dashed lines on the plot delimit three 
regions (E$\in$[1.25,3.00[,[3.00,6.25[ and [6.25,9.75[) and the three 
corresponding $Z_1$ reweighted distributions are plotted in the lower 
left part of the same figure. The middle one, flat and broad, is very 
close to the behavior expected for a critical distribution~\cite{Gul05} and 
illustrates the effect of an energy constraint on the order parameter
distribution. 
If we had made a thinner range, we would have approached the 
microcanonical case. We select the region E$\in$[3.00,6.25[ to compare 
the two reweighted distributions $P_{\omega}^{exp}(E,Z_1)$ and  
$P_{\omega}^{(d.g.)}(E,Z_1)$ (eq.~\ref{Taylor5}). The best solutions 
obtained after this 2D fit procedure are shown in the lower right part 
of figure~\ref{DistriZ1fin} and table~\ref{Resu_Fit2D}. 
They correspond to two ranges of $Z_1$ where fits are performed ([10,79] 
and [25,55]). These two best solutions are shown for the projection on 
the $Z_1$ axis.

\begin{figure}[!hbt]
\begin{center}
\includegraphics[scale=0.80]{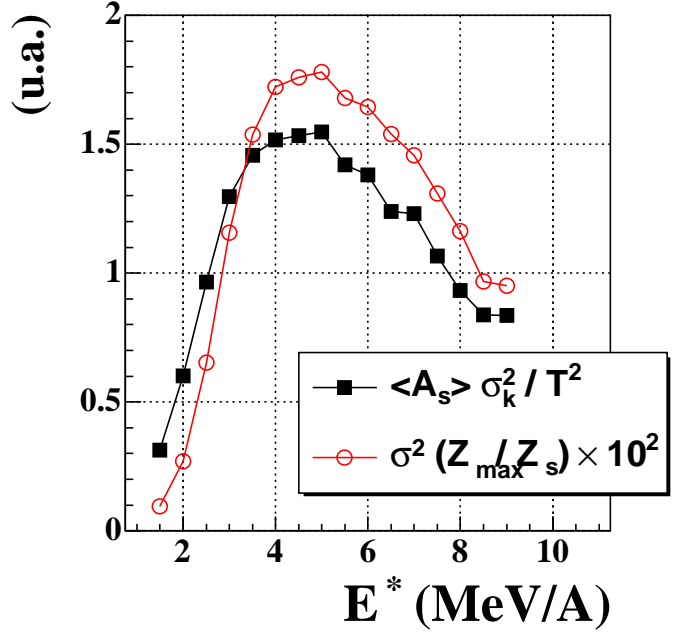}
\caption{\it Microcanonical sample (fixed $E^*$) of the fluctuations 
of normalized FO kinetic energy (open circles) and largest fragment 
charge (full squares). T, As and Zs stand respectively for the 
temperature, the mass and charge of the source.}
\label{fluctu}
\end{center}
\end{figure}

Using two different ranges for the $Z_1$ range allows us to estimate 
the sensitivity of the different parameters. The description of the two 
phases, given by a set of four parameters for each phase, can be summarized
as follows: the average characteristics of the phases, given by 
$\bar{E^{*}}$, $\bar{Z_{1}}$, are well defined. 
The ratio between the populations of the liquid and gas phase strongly 
depends on the interval used to perform the fit. In the two cases the 
normalized estimator, $\chi^{2}$, is good. Concerning the $E^*$ axis, 
the values obtained for the liquid and gas phase centroids reflect the 
location of the coexistence zone, and are consistent with the 
location of the spinodal zone obtained with the AFCE signal with 
the same set of events~\cite{T41Bon06,Ixx-NLN07}. We can further 
explore the coherence between the two signals by looking at the 
fluctuations associated to $Z_1$ and to the Freeze-Out configurational 
kinetic energy~\cite{Cho99}: we observe in fig.~\ref{fluctu} that 
their evolution with excitation energy has a similar behaviour and 
exhibits a maximum for E$\sim$5MeV/A. This observation shows that we 
can consistently characterize the core of the spinodal zone with the 
maximum fluctuations of different observables connected to the order 
parameter of the phase transition.

\section{Conclusion and outlook.}

In this contribution we have shown that, taking into account the dynamics 
of the entrance channel and sorting effects with a renormalization 
procedure, the distribution of the largest size fragment ($Z_1$) of 
each event shows a bimodal pattern. The comparison with an analytical
estimation assuming the presence (the absence) of a phase transition,
shows that the experimental signal can be unambiguously associated 
to the case where the system has a residual convex intruder in its entropy.
This link makes the $Z_1$ observable a reliable order parameter for the 
PT in hot nuclei. A bijective relation between the order of the transition 
and the bimodality signal has been proposed in~\cite{Gul07} and analyses 
on data are in progress.


\end{document}